\input harvmac

\newif\ifdraft\draftfalse
\newif\ifinter\interfalse
\ifdraft\draftmode\else\interfalse\fi
\def\journal#1&#2(#3){\unskip, \sl #1\ \bf #2 \rm(19#3) }
\def\andjournal#1&#2(#3){\sl #1~\bf #2 \rm (19#3) }

\def\ie{{\it i.e.}}
\def\eg{{\it e.g.}}

\def\p{\partial}
\def\ap{\alpha'}

\def\frac#1#2{{#1\over#2}}

\def\half{\frac12}

\def\inbar{\,\vrule height1.5ex width.4pt depth0pt}
\def\IC{\relax\hbox{$\inbar\kern-.3em{\rm C}$}}
\def\IR{\relax{\rm I\kern-.18em R}}
\def\IP{\relax{\rm I\kern-.18em P}}

%
%

\def\prl#1#2#3{Phys. Rev. Lett. {\bf #1} (#2) #3}

\catcode`\@=11
\def\slash#1{\mathord{\mathpalette\c@ncel{#1}}}
\overfullrule=0pt

\def\SS{{\cal S}}

\def\VV{{\cal V}}
\def\WW{{\cal W}}

\def\underrel#1\over#2{\mathrel{\mathop{\kern\z@#1}\limits_{#2}}}

\catcode`\@=12


%

\def\exp{{\rm exp}}


\def\pbar{{\bar \p}}

\def\[{[}
\def\]{]}

\def\comment#1{ }

%
\def\draftnote#1{\ifdraft{\baselineskip2ex
                 \vbox{\kern1em\hrule\hbox{\vrule\kern1em\vbox{\kern1ex
                 \noindent \underbar{NOTE}: #1
             \vskip1ex}\kern1em\vrule}\hrule}}\fi}
\def\internote#1{\ifinter{\baselineskip2ex
                 \vbox{\kern1em\hrule\hbox{\vrule\kern1em\vbox{\kern1ex
                 \noindent \underbar{Internal Note}: #1
             \vskip1ex}\kern1em\vrule}\hrule}}\fi}

%
%






               

%
%
\def\inbar{\hskip.3em\vrule height1.5ex width.4pt depth0pt}
\def\IC{\relax{\inbar\kern-.3em{\rm C}}}
\def\IN{\relax{\rm I\kern-.16em N}}
\def\IQ{\relax\hbox{$\inbar$\kern-.3em{\rm Q}}}
\def\IZ{\relax{\rm Z\kern-.8em Z}}
%
%

%

%
\rightline{EFI-2000-40}
\rightline{RUNHETC-2000-39}
\Title{
\rightline{hep-th/0010108}}
{\vbox{\centerline{ Remarks on Tachyon Condensation}
\vskip 10pt
\centerline{in Superstring Field Theory}}}
\bigskip
\centerline{David Kutasov,$^1$ Marcos Mari\~no,$^2$
and Gregory Moore $^2$}
\bigskip
\centerline{$^1$ {\it Department of Physics, University of Chicago}}
\centerline{\it 5640 S. Ellis Av., Chicago, IL 60637, USA}
\centerline{kutasov@theory.uchicago.edu}
\bigskip
\centerline{$^2$ {\it Department of Physics, Rutgers University}}
\centerline{\it Piscataway, NJ 08855-0849, USA}
\centerline{marcosm, gmoore@physics.rutgers.edu}

\bigskip
\noindent
We generalize recent results on tachyon condensation
in boundary string field theory to the superstring.

\vfill

\Date{October 12, 2000}
 
\lref\sfttach{V. A. Kostelecky and S. Samuel, ``The static
tachyon potential in the open bosonic string theory,'' Phys. Lett. {\bf
B207} (1988) 169;
``On a nonperturbative vacuum for the open bosonic string,''
Nucl. Phys. {\bf B336} (1990) 263;
A. Sen and B. Zwiebach, ``Tachyon condensation in string
field theory,'' hep-th/9912249, JHEP {\bf 0003} (2000) 002;
N. Moeller and W. Taylor, ``Level truncation and
the tachyon in open bosonic string field theory,'' hep-th/0002237,
Nucl. Phys. {\bf B583} (2000) 105.}
\lref\lumps{J.A. Harvey and P. Kraus, ``D-branes as unstable
lumps in bosonic open string theory,'' hep-th/0002117, JHEP {\bf 0004}
(2000) 012; R. de Mello Koch, A. Jevicki, M. Mihailescu, and R. Tatar,
``Lumps and $p$-branes in open string field theory,'' hep-th/0003031,
Phys. Lett. {\bf B482} (2000) 249;
N. Moeller, A. Sen and B. Zwiebach, ``D-branes as tachyon
lumps in string field theory,'' hep-th/0005036, JHEP {\bf 0008}
(2000) 039.}
\lref\liwitten{K. Li and E. Witten,
``Role of short distance behavior in off-shell open-string
field theory,'' hep-th/9303067,
Phys. Rev. {\bf D48} (1993) 853.}
\lref\hkms{J.A. Harvey, S. Kachru, G. Moore, and E. Silverstein,
``Tension is dimension,''  hep-th/9909072, JHEP {\bf 0003} (2000) 001.}
\lref\elitzur{S. Elitzur, E. Rabinovici and G. Sarkissian,
``On least action D-branes,'' hep-th/9807161, Nucl. Phys. {\bf B541}
(1999) 731.}
\lref\gerasimov{A. Gerasimov and S. Shatashvili, ``On exact tachyon
potential
in open string field theory,'' hep-th/0009103.}
\lref\witcs{E. Witten, ``Noncommutative geometry and string field
theory,'' Nucl. Phys. {\bf B268} (1986) 253.}
\lref\lcpp{A. LeClair, M.E. Peskin and C.R. Preitschopf,
``String field theory on the conformal plane, 1. Kinematical
principles,'' Nucl. Phys. {\bf B317} (1989) 411.}
\lref\witbndry{E. Witten, ``On background-independent
open-string field theory,''hep-th/9208027, Phys. Rev. {\bf D46} (1992)
5467. ``Some computations in background-independent
off-shell string theory,'' hep-th/9210065, Phys. Rev. {\bf D47} (1993)
3405.}
\lref\shat{S. Shatashvili, ``Comment on the background independent
open string theory,'' hep-th/9303143, Phys. Lett. {\bf B311} (1993) 83;
``On the problems with background independence in string theory,''
hep-th/9311177.}
\lref\sencon{A. Sen, ``Descent relations among bosonic
D-branes,'' hep-th/9902105, Int. J. Mod. Phys. {\bf A14}
(1999) 4061.}
\lref\witk{E. Witten, ``D-branes and K-theory,'' hep-th/9810188,
JHEP {\bf 9812} (1998) 019.}
\lref\ltsuper{N. Berkovits, ``The tachyon potential in
open Neveu-Schwarz string field theory,'' hep-th/0001084,
JHEP {\bf 0004} (2000) 022; N. Berkovits, A. Sen and
B. Zwiebach, ``Tachyon condensation in superstring field
theory,'' hep-th/0002211; P. De Smet and J. Raeymaekers,
``Level-four approximation to the tachyon potential
in superstring field theory,'' hep-th/0003220, JHEP
{\bf 0005} (2000) 051; A. Iqbal and
A. Naqvi, ``Tachyon condensation on a non-BPS
D-brane,'' hep-th/0004015.}
\lref\berko{N. Berkovits, ``Super-Poincar\'e invariant
superstring field theory,'' hep-th/9503099,
Nucl. Phys. {\bf B450} (1995) 90.}
\lref\hkm{J.A. Harvey, D. Kutasov and E. Martinec,
``On the relevance of tachyons,'' hep-th/0003101.}
\lref\polchinski{J. Polchinski, {\it String Theory}, vol. 2,
Cambridge University Press, 1998.}
\lref\kmm{D. Kutasov, M. Mari\~no, and G. Moore, ``Some
exact results on tachyon condensation in string
field theory,'' hep-th/0009148.}
\lref\goshsen{D. Ghoshal and A. Sen, ``Normalization
of the background independent open string field theory
action,'' hep-th/0009191.}
\lref\mzsuper{J.A. Minahan and B. Zwiebach, ``Effective
tachyon dynamics in superstring field theory,'' hep-th/0009246.}
\lref\supersen{A. Sen, ``Non-BPS states and branes in string
theory,'' hep-th/9904207, and references therein.}
\lref\abs{Atiyah, M. F., Bott, R., Shapiro, A.  ``Clifford modules,''
Topology 3 1964 suppl. 1, 3--38. }
\lref\cornalba{L. Cornalba, ``Tachyon condensation in large magnetic
fields with background independent string field theory,''
hep-th/0010021; K. Okuyama, ``Noncommutative Tachyon from Background
Independent Open String Field Theory,'' hep-th/0010028.}
\lref\hklm{J.A. Harvey, P. Kraus, F. Larsen, and E. Martinec,
``D-branes and Strings as Non-commutative Solitons,''
hep-th/0005031, JHEP {\bf 0007} (2000) 042.}
\lref\wittenk{E. Witten, ``$D$-Branes And $K$-Theory,'' hep-th/9810188,
JHEP {\bf 9812} (1998) 019.}
\lref\horava{P. Horava, ``Type II D-Branes, K-Theory, and Matrix
Theory,'' hep-th/9812135, Adv. Theor. Math. Phys. {\bf 2} (1999) 1373.}
\lref\hori{K. Hori, ``D-branes, T-duality, and Index Theory,'' 
hep-th/9902102}
\lref\hhk{J.A. Harvey, P. Horava, and P. Kraus, ``D-Sphalerons and
the topology of string configuration space,'' hep-th/0001143.}
\lref\lovelace{C. Lovelace, ``Stability of String Vacua. 1. A New
Picture of the Renormalization Group,'' Nucl. Phys. {\bf B273} (1986)
413.}
\lref\hlp{J. Hughes, J. Liu and J. Plochinksi, ``Virasoro-Shapiro
from Wilson,'' Nucl. Phys. {\bf B316} (1989) 15.}
\lref\banksmart{T. Banks and E. Martinec, ``The renormalization
group and string field theory,'' Nucl. Phys. {\bf B294} (1987) 733.}
\lref\atseytlin{A. Tseytlin, ``Sigma model approach
to string theory,'' Int. Jour. Mod. Phys. {\bf A4}
(1989) 1257; ``Renormalization group and string loops,''
Int. Jour. Mod. Phys. {\bf A5} (1990) 589.}
\lref\andreev{O.D. Andreev and A.A. Tseytlin,
``Partition function representation for the open
superstring action,'' Nucl. Phys. {\bf B311} (1988) 205.}
\lref\gms{R. Gopakumar, S. Minwalla, and A. Strominger,
``Noncommutative Solitons,'' hep-th/0003160,
JHEP {\bf 0005} (2000) 020.}
\lref\dmr{K. Dasgupta, S. Mukhi and G. Rajesh,
``Noncommutative Tachyons,'' hep-th/0005006,
JHEP {\bf 0006} (2000) 022.}
\lref\kleb{I. Klebanov and L. Susskind, ``Renormalization group 
and string amplitudes,'' Phys. Lett. {\bf B200} (1988) 446.}
\lref\hkl{J.A. Harvey, P. Kraus and F. Larsen, ``Exact Noncommutative 
Solitons,'' hep-th/0010060.}

\newsec{Introduction}

The boundary String Field Theory (BSFT) of
Witten and Shatashvili \refs{\witbndry,\shat}
is a version of open string field theory in which the
classical configuration space is the space of two dimensional
worldsheet theories on the
disk which are conformal in the interior of the disk
but have arbitrary boundary interactions. Solutions
of the classical equations of motion correspond to
conformal boundary theories. For early work on the closely
related sigma model approach to string theory see \eg\
\refs{\lovelace,\banksmart,\kleb,\andreev,\hlp,\atseytlin}. 

In a recent series of papers   \refs{\gerasimov,\kmm,\goshsen}
it has been  shown that open string tachyon condensation on
D-branes in bosonic string theory can be efficiently
studied in BSFT. In particular:
\item{(a)} Condensation to the closed string vacuum
and to lower dimensional branes involves excitations of
only one mode of the string field -- the tachyon.
\item{(b)} The exact tachyon potential can be computed
in BSFT and its qualitative features agree with Sen's
conjecture \sencon.
\item{(c)} The exact tachyon profiles corresponding
to decay of a higher brane into a lower one give rise
to descent relations between the tensions of various
branes which again agree with those expected from \sencon.

\noindent
In contrast, in Witten's cubic SFT tachyon
condensation in general involves giving expectation
values to an infinite number of components of the
string field. As a consequence, one has to resort
to level truncation \refs{\sfttach,\lumps} and only
approximate results are available.

As explained in \kmm, the reason BSFT gives an efficient
description of tachyon condensation in the bosonic
string is that this process is easy to understand in the
first quantized framework as a property
of the worldsheet renormalization group \hkm. Thus, one
would expect in general that BSFT would give rise to a
useful description of all (classical) physical processes
which correspond to solvable worldsheet RG problems. Exact results 
for tachyon condensation have also been obtained by introducing
noncommutativity \refs{\gms,\dmr,\hklm,\hkl}; the recent results
of \cornalba\ might give a closer relation between this approach 
and BSFT.

In this note we will consider the generalization of the
results of \refs{\gerasimov,\kmm} to tachyon condensation on unstable
brane configurations in the superstring \supersen. There are many
interesting examples of such configurations, both in ten
dimensions and in compactified theories with various degrees
of supersymmetry. There are also some new issues, having to do
with the presence of RR charges carried by some of the branes
that participate in the process of condensation, and the
corresponding spacetime supersymmetry structure. Some aspects
of tachyon condensation in this context have been studied
\ltsuper\ by level truncating the superstring
field theory of \berko.
We will discuss the simplest case --
non-BPS branes and the $Dp-D\bar p$ system in flat ten
dimensional spacetime. As explained in \hkm, the worldsheet
description of condensation is again simple in this case, and
one would expect the BSFT description to be useful.

\newsec{The action in BSFT with worldsheet supersymmetry}

The original papers on BSFT \refs{\witbndry,\shat} studied only
the bosonic case, and as we will see there are some new elements
that arise in the supersymmetric context. Therefore, before
turning to the description of tachyon condensation,
we start with a discussion of BSFT in the superstring.

\lref\aflud{I. Affleck and A. W. Ludwig, ``Universal noninteger
``ground-state degeneracy'' in critical quantum systems,''
\prl{67}{1991}161; ``Exact conformal field theory results on the
multichannel Kondo effect: single fermion Green's function,
self-energy, and resistivity,'' Phys. Rev. {\bf B 48} (1993) 7297.}

Recall that in the bosonic open string, the BSFT action is
constructed as follows. One studies a general worldsheet
theory with boundary interactions, described by the action
\eqn\sso{\SS=\SS_0+\int_0^{2\pi}{d\tau\over2\pi} \VV,}
where $\SS_0$ is a free action defining an open plus closed
conformal background, and $\VV$ is a general boundary perturbation,
which can be parametrized by couplings $\lambda^i$:
\eqn\expv{\VV=\sum_i\lambda^i \VV_i.}
The couplings $\lambda^i$ correspond to fields in spacetime,
and one is interested in constructing the spacetime
action $S(\lambda^i)$. The proposal of \refs{\witbndry,\shat}
is to take the classical spacetime action $S$ to be
\eqn\deftwo{S=(\beta^i{\partial\over\partial\lambda^i}+1)
Z(\lambda),}
where $Z(\lambda^i)$ is the disk partition sum of the worldsheet
theory \sso\ and $\beta^i$ govern the worldsheet RG flow
of the couplings $\lambda^i$ with distance scale $|x|$,
\eqn\rgflow{{d\lambda^i\over d\log|x|}=-\beta^i(\lambda).}
As discussed in \kmm, the action \deftwo\ thus defined
is nothing but the boundary entropy of \aflud.
It coincides with the disk partition sum at RG fixed points,
and decreases along RG flows.

Apriori, one might have expected that the boundary entropy
\aflud\ should be equal to the partition sum
$Z(\lambda^i)$ throughout the RG flow. In the bosonic string,
there are at least two (related) problems with this proposal.
One is the requirement that the boundary entropy should have
critical points whenever the boundary theory is conformal.
This is not necessarily the case for the partition sum
$Z(\lambda^i)$. Indeed,
\eqn\zfd{\partial_i Z=-\int_0^{2\pi}{d\tau\over2\pi}
\langle\VV_i\rangle}
which does not always vanish in a CFT. In particular, for
operators $\VV_i$ of scaling dimension zero there is no
apriori reason for the right hand side of \zfd\ to vanish,
and in general it does not. An example is the
case where $\VV_i$ is taken to be the constant
mode of the tachyon in the bosonic string.

The second problem is that the disk partition sum
is linearly divergent in the bosonic string \atseytlin:
\eqn\divdisk{Z=a_1\Lambda|x|+ a_2.}
$\Lambda$ is a UV cutoff (a large energy); $a_1$ and $a_2$ are finite
coefficients. The origin of the divergence
is the infinite volume of the Mobius group of the disk.

Both problems are avoided by the definition \deftwo.
Indeed, as shown in \refs{\witbndry,\shat}, $S$ can
be alternatively defined by
\eqn\covs{{\partial S\over \partial\lambda^i}=\beta^jG_{ij}(\lambda).}
where $G_{ij}$ is a non-singular metric. Therefore, $S$
is stationary at fixed points of the RG.
Using the Callan-Symanzik equation for
$Z$ one finds that \deftwo\ is equivalent to
\eqn\props{S=Z-{d Z\over d \log|x|}}
in which the divergent term in \divdisk\ precisely cancels.

In the superstring both of the above objections to thinking
of the partition sum as the boundary entropy disappear.
First, worldsheet SUSY implies that all boundary perturbations
\expv\ are top components of worldsheet superfields,
\eqn\topcomp{\VV_i=\{G_{-{1\over2}}, \WW_i\}}
where $\WW_i$ are the bottom components of the corresponding
superfields, and $G_{-{1\over2}}$ is the worldsheet SUSY
generator. The analog of eq. \zfd\ involves in this case the
correlator $\langle \{G_{-{1\over2}}, \WW_i\}\rangle$,
which indeed vanishes at fixed points of the RG, since
in that case $G_{-{1\over2}}$ annihilates both the incoming
and outgoing vacua.

The second objection disappears as well
since  the linear divergence cancels  due to a cancellation
between bosons and fermions -- the supersymmetrically
regularized volume of the super-Mobius group of the disk
is finite \divdisk\ \refs{\atseytlin,\andreev}. 
Note that the above results {\it do not} require
spacetime supersymmetry; they are valid in any vacuum
of the fermionic string.

In view of the above observations, it is natural to propose
that for the superstring, the BSFT action $S$ is simply
the disk partition sum,
\eqn\seqz{S(\lambda^i)=Z(\lambda^i).}
In the context of the low energy effective action for
massless modes this was indeed proposed in \refs{\andreev,
\atseytlin}. We conjecture that this is
the case for the full string field theory in the BSFT
formalism, at least in backgrounds where ghosts and 
matter are decoupled. 

Below, we will use this proposal to study tachyon
condensation in the superstring. Our results can
be viewed as further evidence for the validity of
\seqz. Possible avenues for proving the
conjecture (which we will not attempt here) are:
\item{(a)} The action \deftwo\ was obtained in
\witbndry\ from a Batalin-Vilkovisky (BV) formalism
applied to the space of worldsheet field theories.
It would be interesting to generalize this formalism 
to the fermionic string and derive \seqz.
\item{(b)} A related conjecture is that the disk
partition sum of a supersymmetric worldsheet field
theory coincides with the boundary entropy \aflud.
$Z$ satisfies two of the three properties associated
with the boundary entropy: it is stationary at fixed
points of the RG, and it takes the correct value
there. If one can prove that it decreases along RG
flows, it will be a strong candidate for the boundary
entropy, and thus for the spacetime action in BSFT.

\noindent
Leaving a general derivation of \seqz\ to future work,
we now turn to the example of interest here, tachyon
condensation on unstable D-branes in type II string
theory.\foot{\ie\ Dp-branes with $p\in 2Z+1$ in IIA
string theory, or $p\in 2Z$ in IIB.} We follow closely
the analysis of \refs{\witbndry,\kmm}. The worldsheet
action is
\eqn\actioni{
\SS = \SS_{\rm bulk} + \SS_{\rm boundary}
}
with the standard NSR action in the bulk:
\eqn\bulk{
\SS_{\rm bulk} =
{1\over 4 \pi} \int d^2 z \biggl( \p X^\mu \bar\p X_\mu +
\psi^\mu \pbar \psi_\mu + \tilde \psi^\mu \p \tilde \psi_\mu
\biggr)
}
The integral is over a disk of radius one.
The conventions are those of \polchinski, $\ap = 2$,
and the signature of spacetime is Euclidean.

Supersymmetric boundary interactions corresponding to
open string tachyon condensation are introduced following
the discussion of \hkm. The boundary is described by superspace
coordinates $(\tau, \theta)$, with $0 \leq \tau \leq 2 \pi$ and
$\theta$ the boundary Grassman coordinate.
The boundary superfields  are $\Gamma = \eta + \theta F$ and
${\bf X} = x + \theta \psi$. ${\bf X}$ is the restriction to
the boundary of the standard worldsheet super-coordinate, while
$\Gamma$ is a quantum mechanical degree of freedom which lives
on the boundary. Both $\Gamma$ and ${\bf X}$ are real or, in
the presence of Chan-Paton factors, Hermitian matrices.
The boundary action \actioni\ is given by:
\eqn\boundary{
e^{-\SS_{\rm boundary} } =
{\Tr}P\exp\biggl[ \int {d \tau\over2\pi} d \theta \bigl(
\Gamma D \Gamma + T({\bf X}) \Gamma\bigr) \biggr]
}
where the trace is over the Chan-Paton indices
and $D=\p_{\theta} + \theta \p_{\tau}$. The
fermions $\eta$, $\psi$ are anti-periodic around the
circle (as is appropriate to the NS sector).

Restricting for the moment to the case of one non-BPS brane
(and hence a one-dimensional Chan-Paton space) and performing
the integral over $\theta$, the action \boundary\ becomes
\eqn\bboo{
\SS_{\rm boundary} =-
\int {d \tau\over2\pi} \bigl(
F^2+{\dot\eta}\eta + T(X)F+\psi^\mu\eta\partial_\mu T\bigr)
}
The boundary auxiliary fields are free and can be
easily integrated out. This gives
\eqn\auxflds{\eqalign{F=&-{1\over2} T\cr
\eta=&-{1\over2}{1\over\partial_\tau}(\psi^\mu\partial_\mu T)\cr
& = -{1\over 4} \int d \tau' \epsilon(\tau-\tau') (\psi^\mu\partial_\mu
T)(\tau')\cr
}}
where $\epsilon(x)=+1$ for $x>0$ and $=-1$ for $x<0$.
The formula for $\eta$ in \auxflds\ is non-local but well defined,
since both $\eta$ and $\psi^\mu$ do not have zero modes in
the NS sector. Plugging back into the action \bboo\
one finds
\eqn\bdryinter{\SS_{\rm boundary}= {1\over4}
\int {d \tau\over2\pi} \Biggl[(T(X))^2 +
(\psi^\mu \p_\mu T){1\over\p_\tau}(\psi^\nu \p_\nu T)
\Biggr]
}

\newsec{Free field perturbations}

One of the main points of \kmm\ is that one can learn a lot about
the physics of D-branes by studying solvable boundary perturbations.
In the present context, boundary tachyon profiles of the form
\eqn\btprof{T(X)=a+u_\mu X^\mu}
give rise to free field theory on the worldsheet (see \bboo,
\bdryinter) and can be analyzed by using the results of
\refs{\witbndry,\kmm}. Note that for any non-zero $u_\mu$,
the interaction \btprof\ gives a boundary mass to only one
combination of the superfields ${\bf X^\mu}$. This will play
a role below.

The exact tachyon potential is obtained by setting $u_\mu=0$
in \btprof\ and computing the path integral \bdryinter.
This leads to
\eqn\zeeaa{
S(T) = V_0  e^{-{1\over4} T^2}
}
where $V_0$ is a constant proportional to the volume of the
unstable brane and to $1/g_{\rm string}$. 
This constant  can be normalized by requiring that the
tension of the unstable D-brane, which corresponds to the
vacuum at $T=0$, comes out correctly. This follows from 
\hkms\ and \hkm. One could also perform the consistency check
described in the bosonic case in \goshsen, but we have not
done this.

In any case, the potential term in the
string field theory action on the unstable Dp-brane is
proportional to
\eqn\voft{V(T)= e^{-{1\over4}T^2}.}
This has all the features
expected of the tachyon potential in superstring field
theory: it is symmetric under $T \to - T$, and goes to
zero at $T=\pm\infty$, which corresponds to the closed string
vacuum. In contrast to the bosonic string, the potential
is bounded from below in this case. In \kmm\ it was
proposed that this is related to the absence of a
tachyon in the closed string sector.

To study condensation to lower dimensional branes we next
turn to the case of non-zero $u_\mu$ in \btprof. By a
Poincar\'e transformation we can shift away $a$ and take
$u_\mu$ to point along a single coordinate direction,
$X$. We are thus interested in evaluating the path integral
\bdryinter\ with $T(X)=uX$,
\eqn\pathu{Z(u)=\int [DX][D\psi]e^{-\SS_{\rm bulk}-\SS_{\rm boundary}}}
where
\eqn\ssfree{\SS_{\rm boundary}={u^2\over4}\int_0^{2\pi}
{d\tau\over2\pi}\left(x^2+\psi{1\over\partial_\tau}\psi\right)}
Differentiating with respect to the parameter
\eqn\ydef{y:= u^2}
we have
\eqn\yderv{
{\p \over \p y} \log Z = -{1\over8\pi}
 \int_0^{2\pi} d \tau \langle x^2\ +
 \psi{1\over\partial_\tau}\psi\rangle
}
As in the bosonic case, the correlator that appears in \yderv\
needs to be regularized, since it involves products of fields
evaluated at the same point. In \witbndry\ the correlator
$\langle x^2\rangle$ was defined by point splitting,
\eqn\pointspbos{\langle x^2\rangle=\lim_{\epsilon\to 0}
\left[\langle x(\tau)x(\tau+\epsilon)\rangle-f(\epsilon)\right]}
where $f(\epsilon)$ is a function which has the same
logarithmic singularity as the propagator, so that the
limit \pointspbos\ exists. Of course, this prescription
is ambiguous by a $u$-independent constant (the physical
import of this ambiguity is explained in footnote 2 below).

In the present case, worldsheet supersymmetry leads to
a natural prescription for defining the right hand side
of \yderv:
\eqn\pointspferm{
\langle x^2+\psi{1\over\partial_\tau}\psi\rangle=
\lim_{\epsilon\to 0}
\langle x(\tau)x(\tau+\epsilon)+
\psi(\tau){1\over\partial_\tau}\psi(\tau+\epsilon)\rangle}
To see that this regularization preserves worldsheet supersymmetry
note that by using \auxflds\  the right hand side of
\pointspferm\  is proportional to
\eqn\susypt{\int d\theta\langle {\bf X}(\tau,\theta)
\Gamma(\tau+\epsilon,\theta)\rangle .}
To compute the right hand side of \pointspferm\ we need
the explicit form of the propagators of $x$ and $\psi$
in free massive boundary field theory.
That of $x$ was computed in \witbndry:
\eqn\bosonic{
G_B(\tau-\tau'):=\langle x(\tau) x(\tau') \rangle =
2 \sum_{k \in Z} {1\over\vert k \vert + y} e^{i k (\tau - \tau')}
}
The propagator for fermions on the boundary in the NS sector
is
\eqn\fermionic{G_F(\tau-\tau'):=
\langle \psi(\tau) \psi(\tau') \rangle =
2i  \sum_{k \in Z+\half } {k\over\vert k \vert + y} e^{i k (\tau -
\tau')}
}
To see that \fermionic\ is correct note the following facts.
At $y=0$, it reduces to the familiar result
\eqn\psifree{\langle\psi(\tau)\psi(\tau')\rangle|_{y=0}
=-{2\over\sin{\tau-\tau'\over2}}}
To determine the $y$ dependence consider the
path integral (see \ssfree)
\eqn\propfery{G_F(\tau_1-\tau_2;y)=
\langle \psi(\tau_1) \psi(\tau_2)
e^{-{y\over8\pi}\int_0^{2\pi} d\tau\psi{1\over\partial_\tau}\psi(\tau)}
\rangle}
Differentiating with respect to $y$ we find that
\eqn\diffyy{\partial_yG_F(\tau_1-\tau_2;y)=
-{1\over8\pi}\int_0^{2\pi} d\tau
\langle
\psi(\tau_1) \psi(\tau_2)\psi(\tau){1\over\partial_\tau}\psi(\tau)
\rangle}
Computing the right hand side using the propagator $G_F$ leads
to the differential equation
\eqn\ddffyy{\partial_yG_F(\tau_1-\tau_2;y)=
{1\over4\pi}\int_0^{2\pi} d\tau G_F(\tau_1-\tau;y)
{1\over\partial_\tau}G_F(\tau_2-\tau;y)}
which together with the ``boundary condition'' \psifree\ uniquely
determines the propagator to be \fermionic.

We are now ready to compute the regularized correlator
\pointspferm. Define
\eqn\defgt{\tilde G_F(\epsilon;y):=\langle
\psi(\tau){1\over\partial_\tau}\psi(\tau+\epsilon)
\rangle.}
Comparing to \fermionic\ we see that
\eqn\gteq{\tilde G_F(\epsilon;y)=-2\sum_{k\in Z+{1\over2}}
{1\over |k|+y} e^{ik\epsilon}.}
$\tilde G_F(\epsilon)$ has a very similar form to
$G_B(\epsilon)$ \bosonic,
with the only differences being the overall sign and range
of the index $k$. This is natural, since if the fermions
$\psi$ were periodic, their contribution would precisely
cancel that of the bosons, so the full partition sum would
be trivial \andreev\ -- a consequence of unbroken worldsheet
supersymmetry.

By writing the index $k$ in \gteq\ as one half of an odd
number, and rewriting the sum over odd integers as a difference
of a sum over all integers and that over even integers,
one finds the relation
\eqn\reltgf{\tilde G_F(\epsilon;y)=G_B(\epsilon;y)-
2G_B({\epsilon\over2};2y).}
Substituting in \pointspferm, we find:
\eqn\xxppss{\langle x^2+\psi{1\over\partial_\tau}\psi\rangle
=\lim_{\epsilon\to 0}\left[G_B(\epsilon;y)+
\tilde G_F(\epsilon;y)\right]=
\lim_{\epsilon\to 0}\left[2G_B(\epsilon;y)-
2 G_B({\epsilon\over2};2y)\right].}
To evaluate this limit it is convenient to use the fact
that \witbndry
\eqn\formgg{G_B (\epsilon;y)=-2\log(1-e^{i\epsilon})
-2\log(1-e^{-i\epsilon})+{2\over y} -2y\sum_{k=1}^\infty
{1\over k(k+y)}\left(e^{ik\epsilon}+e^{-ik\epsilon}\right).}
Using this form we find that
\eqn\lineee{\lim_{\epsilon \to 0}
\left[G_B(\epsilon,y)-G_B(\epsilon/2,2y)\right]=
-4 \log 2 +f(y)-f(2y),}
where
\eqn\ffyyii{
f(y)= {2 \over y}-4y \sum_{k=1}^{\infty} {1 \over k(k+y)}}
One can now proceed as in \witbndry\ and integrate the differential
equation \yderv. This leads to
\eqn\finzform{Z(y)= 4^y {Z_1(y)^2 \over Z_1(2y)}}
where  $Z_1$ is a function appearing in the bosonic
case \witbndry,
\eqn\zoney{Z_1(y)=\sqrt{y}e^{\gamma y}\Gamma(y)}
and $\gamma$ is the Euler number. In integrating
the differential equation \yderv\ one also picks up
an integration constant, $Z'$, so the partition sum
is in fact
\eqn\zzoopp{Z=Z' Z(y).}
This is our final result for the partition sum as a function 
of $y$.  The
integration constant $Z'$ can be fixed as discussed above
(after eq. \zeeaa).
One can check that $Z(y)$ \finzform\ is a monotonically
decreasing function of $y$ which approaches the
value 
\eqn\limval{
\lim_{y\to\infty} Z(y)= \sqrt{2\pi}.
}
The result \finzform\ admits several straightforward
generalizations. For example, nontrivial Chan-Paton factors
are easily included by  taking the boundary action
to be
\eqn\bdryintercp{{\Tr} P\exp\Biggl[ -  {1\over 8 \pi}
\int d\tau \biggl((T(x))^2 +
(\psi^\mu \p_\mu T){1\over\p_\tau}(\psi^\nu \p_\nu T)
\biggr) \Biggr]
}
Therefore the potential is proportional to
\eqn\simplpot{
{\Tr} e^{-{1\over4}T^2}
}
Following \refs{\hklm,\cornalba} we can introduce a $B$ field and
take the $B\to \infty$ limit, in which case the action becomes
\simplpot\ with all products now being $*$-products.

Moreover, when there is a nontrivial Chan-Paton space
it is possible to generalize the free-field ansatz in
\btprof\ in an interesting way, which describes condensation
into codimension $M$ branes with $M>1$. Consider a tachyon
profile of the form
\eqn\chanpati{
T(x) = \sum_{i=1}^n  u_i x^i \gamma_i
}
where $\gamma_i$ are Hermitian matrices. Suppose  the matrices
$\gamma_i$
form a Clifford algebra, $\{\gamma_i , \gamma_j\} = 2 \delta_{ij}$.
Since the $\gamma$ matrices are $2^{[n/2]}$ dimensional, the
starting point of the discussion is thus a system of $2^{[n/2]}$
unstable D-branes.

Substituting \chanpati\ into \bdryintercp\ and using the
symmetry of $x^i x^j$ and the Clifford relations we find that
the interaction is proportional to the identity matrix, and
has the form \ssfree\ for the $n$ dimensions $(1,2,\cdots, n)$.
The fact that the interaction is proportional to an identity
matrix in Chan-Paton space means that the $2^{[n/2]}$ unstable
D-branes condense in this case to {\it one} (stable or unstable,
depending on the parity of $n$) codimension $n$ brane. Roughly
speaking, the interaction \chanpati\ acts only on the center
of mass of the $2^{[n/2]}$ unstable branes. The
partition sum \bdryintercp\ is simply equal to
\eqn\newpart{2^{[n/2]}Z'' \prod_{i=1}^n Z(y_i)}
where $y_i=u_i^2$, $Z(y)$ is the function in \finzform, and 
$Z''$ is a constant analogous to $Z'$ above. 

The tachyon profile \chanpati\ is of course the standard
Atiyah-Bott-Shapiro configuration \abs\ which played a
central role in \refs{\wittenk,\horava,\hori}. Restricted to a 
sphere it has unit winding number \abs, which is another 
way to see that the $2^{[n/2]}$ unstable branes are 
condensing to a single brane. It is curious that 
the ABS construction arises in
the present context in a somewhat novel way as a 
configuration preserving the free-field subspace. 

\newsec{Condensation to lower dimensional D-branes}

We can now follow the discussion of \kmm\ to study lower
dimensional
D-branes as solitons in BSFT on a higher dimensional brane.
As a first step, let us determine the spacetime action
\seqz\ resticted to the tachyon field in the 
two-derivative approximation. 
The potential for the tachyon has already been
determined in \voft. By considering the partition sum
$Z$ for general slowly varying tachyon fields $T(X)$ it
is clear that the kinetic term will similarly have
the form $\exp(-T^2/4)\partial_\mu T\partial^\mu T$. To fix
the coefficient we should compare the spacetime action
evaluated with a tachyon profile \btprof\ to the result
\finzform. As in \kmm, this involves regularizing the
volume divergences, as we review next.

Consider, for concreteness, a single non-BPS D9-brane in the
IIA theory. We would like to evaluate the spacetime
action for a profile $T(X^\mu)=uX^1$. It is convenient
to regularize the volume divergence of the remaining
coordinates as in \kmm\ by periodic identification
\eqn\per{
X^{\mu} \sim X^{\mu} + R^{\mu}, \,\,\,\,\,\,\,\, \mu=2, \cdots, 10.}
To determine the correct normalization of the $X$ zero mode,
we notice that, for tachyon profiles of the form $T(X)=uX$,
\eqn\norm{
 \int_{-\infty}^{\infty} {d X \over {\sqrt {2 \pi}}} {\rm e}^{
-{1 \over 4}\int_0^{2 \pi}{d\tau\over2\pi}
(T(X))^2}= \sqrt{2\over y},}
which reproduces the leading term in the expansion
of $Z(y)$ around $y=0$:
\eqn\yzero{
Z(y)=\sqrt{2\over y} + 2 {\sqrt 2} \log 2 \sqrt{y} +
\cdots.}
Therefore, the normalization of the $X$ zero mode is
$1/{\sqrt {2 \pi}}$. The total string field action evaluated
for the boundary perturbation $T(X)=uX^1$ is then given by:
\eqn\stringac{
S(y)=S_0 4^y  {Z_1(y)^2 \over Z_1(2y)}\prod_{\mu=2}^{10}\biggl(
{R^{\mu}  \over {\sqrt {2\pi}}} \biggr).
}
$S_0$ is an overall constant related to $V_0, Z'$ above. 

We can now obtain the exact string field action up to two-derivatives.
It has the form,
\eqn\sfaction{
S=T_9 \int d^{10}x \bigl[ 2\log 2 \, {\rm e}^{- T^2/4}\, \partial^{\mu}
T
\, \partial_{\mu} T + {\rm e}^{- T^2/4} \bigr],}
where the tension of the non-BPS D9-brane is given by
\eqn\tnine{
T_9={S_0 \over (2\pi)^5}.}
As we have discussed above, 
the value of $S_0$ can now be fixed by comparing \tnine\
to the standard tension of a non-BPS D9-brane.
The expression \sfaction\ is obtained by putting
$T(X)=uX^1$ and comparing the kinetic energy with the
second term in the expansion \yzero, and the potential energy
with the first term.\foot{The coefficient of the kinetic term
raises an interesting
technical issue. In the bosonic string case, the choice of
renormalization prescription for $X^2(\tau)$ (see \pointspbos)
renders ambiguous the term in $S$ appearing at order $\sqrt{u}$
in an expansion at small $u$. This has the effect of rendering
ambiguous the coefficient of the two-derivative term for $T$
in the spacetime action.  This ambiguity does not influence
any physical observables, such as the mass of the tachyon
and the tensions of D-branes viewed as solitons. 
In the present case,
the principle of worldsheet supersymmetry removes the
ambiguity (see section 3).} 

The   action \sfaction\ has been recently proposed in
\mzsuper\ as a toy model describing the tachyon dynamics
on a non-BPS D-brane in superstring theory, and conjectured
to be a two-derivative truncation of BSFT.
Notice that the tachyon field configuration $T(X^1)=cX^1$, with
$c=1/(2\log 2)^{1/2}$ is a kink solution of the equations of motion
following from this action:
\eqn\motion{
8\, \log 2\, \partial_{\mu} \partial^{\mu} T - 2 \log 2 \, T
\partial^{\mu} T
\, \partial_{\mu} T + T=0.}
As explained in \supersen, this kink describes a
D8-brane of type IIA theory and one can
compute the tension ratio $T_8/T_9$ by plugging 
the kink profile in \sfaction. Since
this configuration is not a solution of the equations
of motion of the full string field action (which includes an
infinite number of higher derivative terms),
one only gets in this way an approximate value for the
tension of the D8-brane.

However, as in the bosonic case \kmm, we know that the exact
profile of the soliton in the exact string field theory will also
be of the form $T(X)=uX$ for some value of $u$. The reason
is that this particular tachyon mode corresponds to a free field
theory on the worldsheet, and does not mix with the rest of the
modes. Therefore, to obtain the exact profile we just have to
take $u$ to correspond to the infrared attractive fixed point of the
RG flow \hkm. This is the value of $u$ that minimizes the string field
action \stringac. Since \finzform\ is
monotonically decreasing, the minimum is achieved at
$y_*= \infty$, as expected from the RG flow
picture. At this infrared fixed point, the exact value
of the action is:
\eqn\sinfty{
S(y_*)=S_0{\sqrt {2 \pi}} \prod_{\mu=2}^{10} \biggl( {R^{\mu}   \over
{\sqrt
{2\pi}}} \biggr).}
We can now determine the ratio of the D-brane tensions.
{}From the spacetime point of view, the kink 
describes a D8-brane, therefore we have $S(y_*)=
T_8 \prod_{\mu}R^{\mu}$. After restoring units, 
taking into account the fact
that $\alpha'=2$, we conclude that
\eqn\ratio{
T_8 =(2 \pi \sqrt{\ap}) {T_9 \over \sqrt{2}} ,}
which is the expected value:\foot{Using the exact action
up to two derivatives \sfaction, one finds a ratio of tensions
which is $1.151$ times the expected value (see also \mzsuper).}
the tensions of
BPS Dp-branes are proportional to
$(2\pi \sqrt{\ap})^{3-p}/\kappa_{10}$ \polchinski. The tensions
of the non-BPS branes are larger by a factor of $\sqrt{2}$.

In contrast to the bosonic case, the D8-brane obtained
in this way is stable, and cannot further decay.
This is consistent with the fact that the tachyonic
profile which gives a free theory on the
worldsheet is {\it linear} in $X$: as we remarked above, by
a Poincar\'e transformation we can always take it
to be of the form $uX^1$, and therefore it always
describes a codimension one kink. At first sight
this appears to be a problem for describing tachyon
condensation to lower dimensional branes since
one can only condense a single direction in spacetime.
However, as discussed above, adding Chan-Paton factors and
taking a tachyon profile as in \chanpati\ leads to condensation
of $2^{[n/2]}$ $D9$-branes to a single $D(9-n)$-brane.
Repeating the considerations of this section for the
partition sum \newpart\ we conclude that BSFT gives rise
to the descent relation
\eqn\highd{{T_{9-n}\over2^{[n/2]} T_9}=(\pi\sqrt{2\alpha'})^n,}
or, equivalently
\eqn\rightratios{
{T_{9-n}\over T_9} =
\cases{
(2\pi \sqrt{\ap})^n &\qquad\qquad n ~~ {\rm even}\cr
{1\over\sqrt{2}}(2\pi \sqrt{\ap})^n  &\qquad\qquad n ~~ {\rm odd}\cr}
}
which is indeed the correct answer.\foot{Note that the above
discussion does {\it not} imply that a single unstable D9-brane
cannot decay to Dp-branes with $p<8$. As explained in \hkm\ it
can, but not within the free field subspace of configuration
space which we are focusing on here.}

One of the most important attributes of type II
D-branes is that they carry RR charge \polchinski. Since
BSFT is so closely allied to worldsheet techniques
this property is easily checked in the present
formalism using standard techniques. We follow 
a computation described in \wittenk. To compute the RR charge
we compute the one-point function on the disk of the
RR vertex operator in the $(-3/2,-1/2)$ picture,
\eqn\vrrr{
V = C_{\dot a \dot b} S^{\dot a}(z) \tilde S^{\dot b}(\bar z)
e^{-3/2 \phi(z) - 1/2 \tilde \phi(\bar z) } + \cdots.}
Here $\dot a, \dot b$ are chiral spinor indices for $SO(10)$,
$C$ is the RR potential,   $S^{\dot a}$ are spin operators,
and $\phi,\tilde\phi$ are bosonized superconformal ghosts.
When such a vertex operator is inserted into the disk,
$\eta, \psi$ become periodic. In particular, we must soak up
the $\eta, \psi $ zero-modes. The tachyon vertex operator is
\eqn\tachvv{V_T = \int {d \tau \over 2\pi} d \theta \Gamma T(X) =
\int {d \tau \over 2\pi} (FT(x)+ \eta \psi^\mu\partial_\mu T(x)).}
The resulting spacetime interaction is proportional to
$\int C \wedge e^{-{1\over 4}T^2} d T$, and, in a solitonic
background $T= u X$, the charge is proportional
to $u/\vert u\vert$. Note that the charge density is distributed
equally on both sides of the D8 brane in accord with \hhk.

Finally, the above discussion has focused on the case of
an unstable D-brane, but it can be extended to $D\bar D$ systems
by considering tachyon configurations of the form
$\pmatrix{0 & T \cr T^\dagger & 0 \cr}$.

In this paper we have made some preliminary remarks
on the formulation of BSFT in the
superstring case.  Our results indicate that this
approach might be a useful alternate
route to superstring field theory.
In addition to the open problems listed in \kmm, the
supersymmetric case raises many further interesting
open questions. Further development of BSFT for the
superstring appears to be a worthwhile enterprise.

\bigskip
\noindent{\bf Acknowledgements:}
We would like to thank T. Banks, H. Liu, E. Martinec,
A. Rajaraman, M. Rozali and A. Tseytlin
for useful discussions. The work of DK is supported
in part by DOE grant \#DE-FG02-90ER40560. The work
of GM and MM is supported by DOE grant DE-FG02-96ER40949.
MM would like to thank the High Energy Theory group at
Harvard for hospitality in the final stages of this work.

\listrefs

\bye